%% file: conf_paper.tex
\documentclass[11pt]{article}
\usepackage{graphicx}

\newcommand{\BABARPubYear}    {02}

\newcommand{\BABARConfNumber} {04}
\newcommand{\SLACPubNumber} {9171}

\input babarsym

\long\def\inst#1{\par\nobreak\kern 4pt\nobreak
    {\it #1}\par\vskip 10pt plus 3pt minus 3pt}

\begin{document}
{\pagestyle{empty}

\begin{flushright}
\babar-CONF-\BABARPubYear/\BABARConfNumber \\
SLAC-PUB-\SLACPubNumber \\
\end{flushright}

\par\vskip 5cm

\begin{center}
\Large \bf A Measurement of the $\Bz \to \jpsi\pi^+\pi^-$ Branching Fraction
\end{center}
\bigskip

\begin{center}
\large The \babar\ Collaboration\\
\end{center}
\bigskip \bigskip

\thispagestyle{empty}
\begin{center}
\large \bf Abstract
\end{center}
We present a preliminary measurement of the branching fraction for the decay of
the neutral \B\ meson into the final state $\jpsi\pipi$.  The data set
contains approximately 56 million \BB\ pairs produced at the \FourS\ 
resonance and recorded by the \babar\ detector at the \pep2\ \epem\ collider. 
The result of this analysis is $\mathcal{B}$($\Bz\to\jpsi\pipi$) = 
(5.0 $\pm$ 0.7 $\pm$ 0.6)$\times 10^{-5}$, where the first error is 
statistical and the second is systematic.

\vfill
\begin{center}
Presented at the XXXVII$^{th}$ Rencontres de Moriond on
QCD and Hadronic Interactions,\\
3-16---3/23/2002, Les Arcs, Savoie, France
\end{center}

\vspace{1.0cm}
\begin{center}
{\em Stanford Linear Accelerator Center, Stanford University, 
Stanford, CA 94309} \\ \vspace{0.1cm}\hrule\vspace{0.1cm}
Work supported in part by Department of Energy contract DE-AC03-76SF00515.
\end{center}
}

\newpage

\input authors_win2002

\section{Introduction}
\label{sec:Introduction}
In this paper, we report a preliminary measurement of the $\Bz\to\jpsi\pipi$
branching fraction.  The motivation for studying these decays comes in part 
from the possibility of including the $\jpsi\rho^0$ mode in the \babar\ 
\stwob\ analysis~\cite{Aubert:2001nu}.  It is also expected that due to the 
Cabibbo and colour suppression of these decays, interference with rare or 
exotic processes, such as a box diagram containing charged Higgs bosons, could
be significant.  These effects may appear as deviations of the branching ratio 
from the Standard Model prediction of 
$\mathcal{B}(\Bz\to\jpsi\pipi)=(4.6\pm 0.8)\times 10^{-5}$~\cite{Bishai:1995yj}.  The only previous measurement related to this analysis is an upper limit on
the $\Bz\to\jpsi\rho^0$ branching fraction by CLEO~\cite{Bishai:1995yj}, 
$\mathcal{B}(\Bz\to\jpsi\rho^0) < 2.5\times 10^{-4}$ at 90\% confidence level.

The method used is a fit to the distribution of the 
invariant mass of the two pions, $M(\pipi)$, with the goal of isolating the
separate components.  Both the $\Bz\to\jpsi\rho^0$ and $\Bz\to\jpsi\pipi$ 
(non-resonant) contributions are modelled in the fit function.  However, in 
the present analysis the branching fraction is quoted for the sum of these two
modes, as the individual fractions are highly correlated and depend 
sensitively on model assumptions. The total yield is only sensitive to the 
normalization of various background sources and so is quite stable.
It is expected that separation of the two 
components will be possible with a considerably larger data set, when a mass 
dependent angular analysis of the decay products will be possible.

\section{The \babar\ detector and data set}
\label{sec:babar}
The data used in this analysis were collected with the \babar\ detector
at the \pep2\ storage ring.  The \babar\ detector is described in detail
elsewhere~\cite{ref:babar}.  Surrounding the beam pipe is a silicon vertex
tracker (SVT) to provide precise measurements of positions and angles of 
charged particles emerging from the interaction region and stand-alone track 
finding, particularly for particles with momentum below 120\,\mevc.  Outside 
this is a 40-layer drift chamber (DCH) filled with an 80:20 helium-isobutane 
gas mixture to minimize multiple scattering.  The DCH provides measurements of
track momenta, as well as energy-loss measurements that contribute to charged 
particle identification.  Surrounding the drift chamber is a novel detector of 
internally reflected Cherenkov radiation (DIRC) that allows charged hadron 
identification in the barrel region.  Outside the DIRC is a CsI(Tl) 
electromagnetic calorimeter
(EMC), which is used to detect photons and neutral hadrons and to provide 
electron identification.  The EMC is surrounded by a superconducting coil, 
which provides a magnetic field of 1.5T for momentum measurements.  Outside 
the coil the flux return is instrumented with resistive plate chambers 
interspersed with iron (IFR) for the identification of muons and long-lived 
neutral hadrons.  The GEANT4~\cite{geant4} software is used to simulate 
interactions of particles traversing the \babar\ detector.  A coordinate system
is defined with the $z$ axis aligned along that of the detector in the 
electron beam direction.

The data sample used for the analysis contains approximately 56 million \BB\
pairs, corresponding to 51.7\invfb\ taken near the \FourS\ resonance.  An 
additional 6.4\invfb\ of data, taken approximately 40\mev\ below the 
\FourS\ peak, was used in studies of the effect of light \qqbar\ pair 
backgrounds.

\section{Analysis method}
\label{sec:Analysis}
\subsection{Event selection and $B$ reconstruction}
\label{sec:selection}
The aim of the event selection is primarily to accept events containing a 
$\Bz\to\jpsi\pipi$ decay while rejecting background both from $u,d,s,c$ quark 
continuum and $\FourS$ events which do not contain a $\Bz\to\jpsi\pipi$ decay.

Events containing \BB\ pairs are selected based on track multiplicity and 
event topology.  Tracks in the polar angle region 
$0.41 < \theta_{\rm lab} < 2.54$\,\rad, with transverse momentum 
greater than 100\mevc\, are required to pass quality cuts, including number of 
DCH hits used in the track fit and impact parameters with respect to the 
nominal beam spot in the $r-\phi$ and $r-z$ planes.  At least three tracks 
must pass the above selection.  To reduce continuum background the ratio of 
second to zeroth Fox-Wolfram moments~\cite{Fox:1978vw}, $R_2 = H_2/H_0$, is 
required to be less than 0.5.  The total charged and neutral energy must be 
greater than 4.5\gev\ in the fiducial region of the detector (charged tracks 
in the DCH and neutral clusters in the EMC) and the primary vertex of the 
event must be within 0.5\cm\ of the average measured beam spot position in the
plane transverse to the beamline.  

Electron candidates must satisfy the requirement that the ratio of calorimeter
energy to track momentum lies in the range $0.75 < E/p < 1.3$, have a 
cluster shape and size consistent with an electromagnetic shower, and DCH 
\dedx\ consistent with an electron.  If an EMC cluster close to the electron 
track is consistent with originating from a bremsstrahlung photon, it is added 
to the electron candidate.   

Muon candidates must satisfy requirements on the number of interaction lengths
of IFR iron penetrated of $N_{\lambda} > 2$, the difference in the 
measured and expected interaction lengths penetrated of 
$|N_{\lambda} - N^{exp}_{\lambda}| < 2$, the position match between the
extrapolated DCH track and the IFR hits, and the average and spread of the 
number of IFR strips hit per layer.  

Pion candidates are accepted if they originate from close to the beam spot and
fail a charged kaon identification algorithm that is designed to reject 
pions.  The algorithm uses \dedx\ information from the SVT and DCH, and 
Cherenkov angle and number of photons from the DIRC.  

Tracks are required to lie in fiducial ranges within which the efficiency is
well known from control samples and the material in the detector is accurately
modelled in the simulation.  The accepted ranges in laboratory polar angle are
$0.41 < \theta_{\rm lab} < 2.41$\,\rad for electrons, 
$0.3 < \theta_{\rm lab} < 2.7$\,\rad for muons, and 
$0.35 < \theta_{\rm lab} < 2.5$\,\rad for pions.  These ranges correspond 
approximately to the geometrical acceptances of the EMC, IFR and DIRC, 
respectively.

$\jpsi\to\epem$ or $\mumu$ candidates are formed from pairs of
identified electrons or muons which are fitted to a common vertex and must
lie in the invariant mass interval 2.95 (3.06) to 3.14\gevcc\ for the 
\epem\ (\mumu) channel.

\Bz\ candidates are formed by combining a $\jpsi$ candidate with pairs of oppositely-charged pion candidates that are fitted to a common vertex.  We require 
that both two-prong vertices coincide spatially by demanding
\begin{equation}
\chi^2_{\psi,\pi\pi} = 
\Sigma_i \frac{(x^i_{\psi}-x^i_{\pi\pi})^2}{(\sigma(x^i_{\psi})^2 + 
\sigma(x^i_{\pi\pi})^2)} < 6,
\end{equation}
where $x^i$ and $\sigma(x^i)$ are the positions and errors of
the vertices in the $i$th spatial dimension.  Further selection requirements 
for $B^0$ candidates are made using two kinematical variables: the 
difference $\Delta E$
between the energy of the candidate and the beam energy $E^{\rm cm}_{\rm beam}$
in the center-of-mass frame and the beam-energy substituted mass 
$m_{\rm ES} = \sqrt{(E^{\rm cm}_{\rm beam})^2 - (p^{\rm cm}_B)^2}$.  After
applying the loose requirements $5.2 < m_{\rm ES} < 5.3\gevcc$ and 
$| \Delta E | < 120\mev$, if more than one \Bz\ candidate remains in the same 
event only the candidate with the smallest $| \Delta E |$ is kept.  Those 
candidates with $| m_{\rm ES} - 5279.0 | < 9.9\mevcc$ and 
$| \Delta E | < 39\mev$ form the final signal sample.  These ranges correspond
to $3\sigma$ in the expected resolutions for $m_{\rm ES}$ and $\Delta E$.  
After all selection criteria have been applied, there are 212 events 
remaining.

\subsection{Fitting to the \pipi\ mass spectrum}
An unbinned, extended maximum likelihood fit is performed on the invariant 
mass distribution of the two pions for selected events.  We consider five 
categories of events:

\begin{itemize}
\item $\Bz\to\jpsi\pipi$ (non-resonant) events
\item $\Bz\to\jpsi\rho^0$ events
\item $\Bz\to\jpsi\KS (\KS\to\pipi)$ events
\item Background from events with a fake $\jpsi$
\item Inclusive-$\jpsi$ background (that originating from events containing a 
real $\jpsi$)
\end{itemize}
A probability density function (PDF) is constructed for each of these five 
cases.  The total PDF is then formed from the sum of the five PDFs and fitted 
to the data.  The $\Bz\to\jpsi\KS$ mode is not considered signal 
for the purposes of determining the branching fraction $\Bz\to\jpsi\pipi$.  

\subsubsection{The signal and \KS\ PDFs}
\label{sec:pdfs1}
The PDF used to model the $\Bz\to\jpsi\pipi$ (non-resonant) mode is the 
two-body phase space distribution of the pion pair and is given by
\begin{equation}
F_{n.r.}(m) = (m-M_{on})^A \times (M_{off}-m)^B,
\end{equation}
where the kinematic limits are $M_{on} = 2M_{\pi}$ and $M_{off} = 
M_{B^0} - M_{\jpsi}$.  The parameters $A,B$ are determined by 
fitting to the $M(\pipi)$ distribution in simulated $\Bz\to\jpsi\pipi$
(non-resonant) events.

The PDF used to model the $\Bz\to\jpsi\rho^0$ mode is a relativistic p-wave
Breit-Wigner~\cite{Jackson:zd} function multiplied by the phase-space shape 
described above.  The $\rho^0$ mass and width for the Breit-Wigner 
are fixed to their PDG values~\cite{Groom:in}.  

The $\Bz\to\jpsi\KS$ two-pion invariant mass distribution is modelled by a 
single Gaussian function with the mass and width fixed to values obtained by 
fitting a sample of simulated $\jpsi\KS$ events.  

\subsubsection{The background PDFs}
\label{sec:pdfs2}
The $M(\pipi)$ distribution for events without a real $\jpsi$ is derived from 
a fake-$\jpsi$ sample from data.  The sample is selected as in 
Sec.~\ref{sec:selection} except that at least one of the lepton candidates 
must fail the appropriate particle identification requirements, thus providing 
a high statistics data set that 
models the shape of the non-$\jpsi$ background.  In addition the final 
kinematic requirements are relaxed to $5.2 < m_{\rm ES} < 5.3\gevcc$ and 
$| \Delta E | < 120\mev$, which further enlarges the sample.  A Monte Carlo 
study confirms that the $M(\pipi)$ distribution obtained with this procedure
correctly describes the shape of the non-$\jpsi$ background.  
The resulting distribution is parameterized using the sum of two Weibull 
functions (see Eq.~\ref{eqn:weibull}) and a Breit-Wigner.
 
The $M(\pipi)$ background shape from events containing a real $\jpsi$ is
obtained with a sample of simulated $\B\to\jpsi\X$ events equivalent to a 
luminosity of 81\invfb.  Events where the system $X$ is \pipi (non-resonant), 
$\rho^0$ or \KS(\pipi)\ are removed from the 
sample.  The resulting shape is well described by a Weibull 
function~\cite{Weibull}:
\begin{eqnarray}
\label{eqn:weibull}
F_{J/\psi X}(m) & = & C\,V\,(m-M_{on})^{(C-1)} \times \exp 
[-V (m-M_{max})^C] \nonumber \\
V & = & (C-1)/(C(M_{max}-M_{on})^C),
\end{eqnarray}
where $M_{max}$ (the peak position) and $C$ are determined from the fit to
the simulated events and $M_{on}$ is the kinematic limit as described 
above.

\subsubsection{Background normalization}
The normalization of the background components is obtained from control 
samples in data.  The level of non-$\jpsi$ background is obtained from the
data $\jpsi$-sideband sample.  This sample is selected by accepting only 
events 
where the $\jpsi\to\epem$ candidate lies in the invariant mass region 3.156 to 
3.300\gevcc\ or the $\jpsi\to\mumu$ candidate lies in the region 2.980 to 
3.024\gevcc\ or 3.156 to 3.300\gevcc.  The $m_{ES}$ distribution is fit to an 
ARGUS function~\cite{Albrecht:1990cs} for $B$ candidates in which the 
$\jpsi\to\epem$ or $\jpsi\to\mumu$ candidate lies in the $\jpsi$ sideband.  
The fit is done separately for $\epem$ and $\mumu$ candidates to obtain the 
number of background events passing the final kinematic selection in each 
case.  These numbers are then reweighted to make them correspond to the fitted
level of background in the $M(\epem)$ or $M(\mumu)$ distribution in the 
$\jpsi$ mass interval.

The normalization of inclusive-$\jpsi$ background is obtained from the 
distribution of $m_{ES}$ for events in the $\Delta E$ signal region.  The 
$m_{ES}$ distribution is parameterized by a Gaussian (to represent signal and 
peaking background) and an ARGUS function, which has one shape parameter and 
an endpoint fixed by the average beam energy in the center-of-mass frame.  
The peaking background is defined as that which accumulates at 
$m_{ES}=5.279\,\gevcc$.  It originates from $\B\to\jpsi\X$ decays such as 
$\B\to\jpsi\Kstar$.  The non-peaking component of the inclusive-$\jpsi$ 
background comes from subtracting the non-$\jpsi$ contribution, on the basis 
of the scaled sideband events described above, from the total ARGUS background
in data.  We then add a peaking component, which is obtained from the Gaussian
part of the $m_{ES}$ distribution in $\B\to\jpsi\X$ simulation (with $X=\pipi$ 
(non-resonant), $\rho^0$ and $\KS (\pi^+\pi^-)$ removed).  The total 
background from events containing a real $\jpsi$ is then the sum of peaking 
and non-peaking components.  
The peaking component comprises only 5\% of the total background.  Thus any 
associated uncertainties, such as those on the branching ratios used in the 
$\jpsi X$ simulation, will have a relatively small effect on 
the final systematic error.

\subsection{Branching fraction determination}
The branching fraction is obtained from
\begin{equation}
\label{eqn:bf}
\mathcal{B}(\Bz\to\jpsi\pi^+\pi^-) = 
\frac{N_{\pi\pi}}{N_{\Bz}\times\epsilon_{\pi\pi}\times\mathcal{B}
(\jpsi\to\ellell)},
\end{equation}
where $N_{\pi\pi}$ is the yield of resonant and non-resonant components 
obtained from the fit, $N_{\Bz}$ is the total number of \Bz and \Bzb in the 
data sample~\cite{Aubert:2001xs} and $\epsilon_{\pi\pi}$ is the signal 
efficiency.  The $\jpsi$ branching fraction $\mathcal{B}(\jpsi\to\ellell)$ is 
fixed to the current world average value~\cite{Groom:in}.  We assume
that the branching fraction $\FourS\to\Bz\Bzb$ is 50\%.

The signal efficiency in Eq.~\ref{eqn:bf} is derived from simulated signal
events.  Imperfect modelling of particle identification is corrected by 
studies of independent control samples derived from data.  
The lepton and pion particle identification efficiencies are measured in data
with control samples of $\mumu\g$, $\mumu\epem$, \epem, $\epem\g$, 
$\Dstarp\to\Dz\pip$ ($\Dz\to\Km\pip$) and \KS\to\pip\pim.  The efficiencies 
are determined as a function of momentum, and polar and azimuthal angle.  We 
find $\epsilon(\jpsi\rho^0)=(27.0\pm0.3)\%$, $\epsilon(\jpsi\pipi,$ 
non-resonant$)=(26.5\pm0.3)\%$ and $\epsilon(\jpsi\KS)=(2.31\pm0.06)\%$.
The final corrected signal efficiency is taken as the average of the 
$\jpsi\rho^0$ and $\jpsi\pipi$ (non-resonant) efficiencies and is found to be 
($26.8 \pm 0.2$)\%, where the error is from Monte Carlo statistics.  

\section{Results}
\label{sec:Physics}
\begin{figure}[htbp]
\begin{center}
\includegraphics[height=10cm]{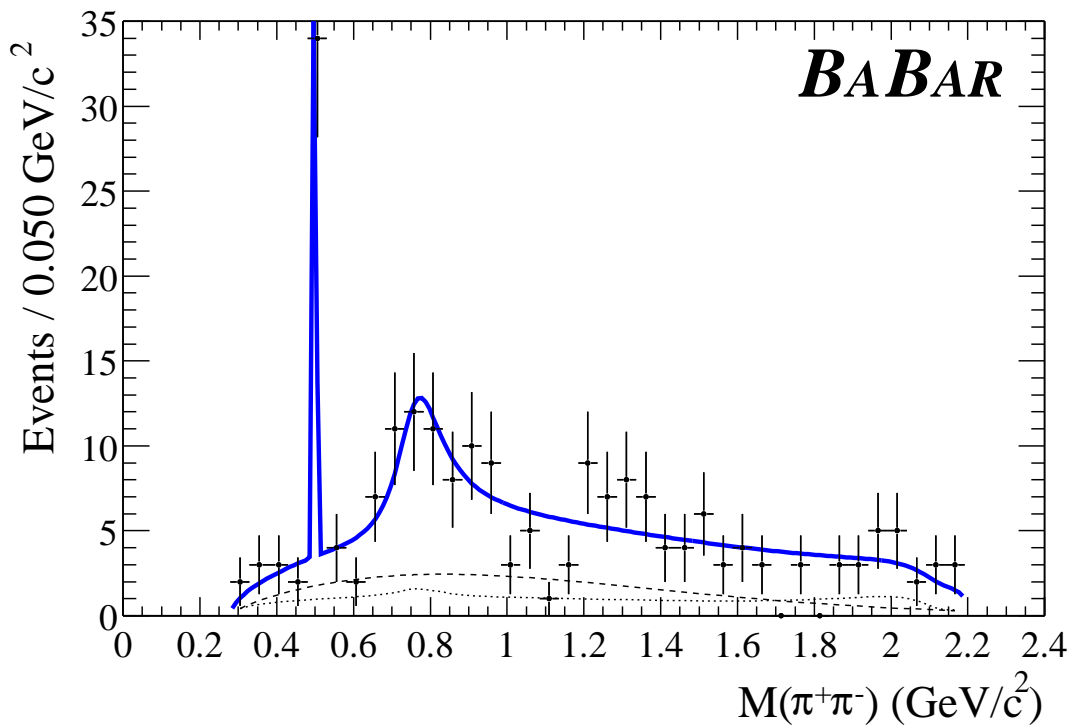}
\vspace*{0cm}
\caption{Distribution of the invariant mass M(\pipi) for events passing all 
selection criteria. The solid line is the result of the unbinned likelihood 
fit. The dotted (dashed) line represents the background from non-$\jpsi$
(inclusive-$\jpsi$) events.}
\label{fig:fit}
\end{center}
\end{figure}

The full model for the $M(\pipi)$ mass distribution is obtained by summing the
five PDFs described in sections~\ref{sec:pdfs1} and~\ref{sec:pdfs2}.
A likelihood fit is performed on the
$M(\pipi)$ distribution in data with the normalization of the non-$\jpsi$ 
background fixed to 36.9 events and the inclusive-$\jpsi$ background
to 55.1.  Thus the only parameters that are allowed to float in the fit are 
the numbers of $\jpsi\pipi$ (non-resonant), $\jpsi\rho^0$ and $\jpsi\KS$
events.  The result of the fit is shown in Fig.~\ref{fig:fit} where the data 
has been binned and overlayed.  The $\chi^2$ of the curve and the data as 
binned in Fig.~\ref{fig:fit} is 40.9 for 38 data points.
The fit finds $N_{\pi\pi} = 89.9 \pm 13.3$ events, where $43.3\pm12.7$ are 
included in the $\rho^0$ component and $46.6\pm14.5$ are included in the 
non-resonant component.  The number of events in the $\KS$ component is 
$29.0\pm5.5$.  Inserting the result into
Eq.~\ref{eqn:bf} yields the branching ratio
\begin{displaymath}
\mathcal{B}(\Bz\to\jpsi\pipi) = (5.0 \pm 0.7)\times 10^{-5},
\end{displaymath}
where the error is statistical.  The results are summarised in 
Table~\ref{tab:res}.

\begin{table}
\caption{Number of events for each signal and background category floated in
the fit.  $\jpsi\pipi$(counting) is the $\jpsi\pipi$
yield determined by counting the number of events that pass all selection
criteria and subtracting the background and $\KS$ estimates from it, as 
described in Sec.~\ref{checks}.}
\begin{center}
\begin{tabular}{|c|c|}
\hline
$B$ decay mode & Yield \\ \hline \hline
$\jpsi\rho^0$ & $43\pm13$ \\ \hline
$\jpsi\pipi$ (non-resonant) & $47\pm15$ \\ \hline
$\jpsi\KS (\KS\to\pipi)$ & $29\pm6$ \\ \hline
$\jpsi\pipi$ (fit) & $90 \pm 13$ \\ \hline \hline
$\jpsi\pipi$ (counting) & $91 \pm 17$ \\ \hline
\end{tabular}
\label{tab:res}
\end{center}
\end{table}

\section{Systematics studies}
\label{sec:Systematics}
The systematic errors on the final branching fraction measurement arise from
uncertainties in the signal efficiency, fitted yield, number of \BB\ pairs
produced and $\jpsi\to\ellell$ branching fraction.  $N_{\BB}$ is known to 
$1.1\%$ with the dominant contribution to the uncertainty coming from the error
on the efficiency of the \B-counting selection.  
$\mathcal{B}(\jpsi\to\ellell)$ is known to 1.7\%~\cite{Groom:in}.

The uncertainty on the pion particle identification efficiency is 2.5\%
per pion.  This is determined in part by the limited number of events in the 
control sample.
Other contributions are from applying different techniques to deduce the kaon 
contamination in the control sample and deviations between the efficiency for 
pions in the control sample and in the $\jpsi\pipi$ sample.  Thus the final 
systematic error from pion identification is 5\%.  Uncertainties on electron 
and muon particle identification efficiencies come from studies of yields in 
$B\to\jpsi X$ data events and give an error of 1.3\%.  The total 
systematic error on particle identification efficiency is then 5.2\%.

The uncertainty in the determination of the tracking efficiency is 1.3\% per 
track, which sums coherently for the four tracks coming from the \Bz.  

The efficiency of the convergence requirement on the $\pipi$ vertex fit has 
been studied with a sample of $\psitwos\to\ellell$ decays. Data and simulation
are found to be in good agreement with an associated systematic error of 1\%.

The $\jpsi\rho^0$ component of the final sample will introduce an uncertainty
on efficiency due to the unknown polarisation of the $\rho$.  Studies of 
efficiency variations between samples of simulated events with longitudinal 
and transverse polarisations lead to a systematic error estimate of 2.5\%.

To account for imperfect simulation of selection variables, the requirements 
on $\jpsi$ invariant mass, $\chi^2_{\psi,\pi\pi}$
and $|\Delta E|$ are varied within reasonable limits, and the analysis is 
repeated for each selection.  The deviation of the efficiency-corrected yield 
from the fit, in each case, is assigned as a systematic error.  Varying the 
size of the $\jpsi$ mass interval by $\pm 14\mevcc$ and the $|\Delta E|$ 
requirement by $\pm 1\sigma$, where $\sigma=13\mev$ is the standard deviation 
observed in data, shows variation at the level of 0.4\% and 6.8\%, 
respectively.
The $\chi^2_{\psi,\pi\pi}$ requirement is varied between 3 and 9 yielding an 
error of 2.1\%.  These errors are added in quadrature to give the total 
systematic uncertainty due to modelling of the selection requirements.

To assess the effect of the chosen signal and background shapes on the fitted 
yields, the fixed parameters of these PDFs are varied within their statistical
errors, allowing for correlations.  This produces a total systematic error due
to fit parameter variation of 4.9\%.  The total systematic uncertainty from 
all sources is found to be 11.8\%, as summarized in Table~\ref{tab:syst}.

\begin{table}
\caption{Summary of the systematic errors for $\mathcal{B}(\Bz\to\jpsi\pipi$).}
\begin{center}
\begin{tabular}{|c|c|}
\hline 
Source of Uncertainty & Systematic Error \\
\hline
\hline
B-Counting & 1.1\% \\
\hline
$\sigma(\mathcal{B}(\jpsi\to\ellell))$ & 1.7\% \\
\hline
PID efficiencies & 5.2\% \\
\hline
Track efficiencies & 5.2\% \\
\hline
$\pi\pi$ vertex efficiency & 1.0\% \\
\hline
$\rho^0$ polarisation & 2.5\% \\
\hline
Selection variation & 7.1\% \\
\hline
PDF parameter variation & 4.9\% \\
\hline
Simulation statistics & 0.7\% \\
\hline
\hline
Total & 11.8\% \\
\hline
\end{tabular}
\label{tab:syst}
\end{center}
\end{table}

\section{Cross checks}
\label{checks}
The fit was repeated using a simple Breit-Wigner (non-relativistic, width 
independent of mass) for the $\rho^0$ lineshape.  While the $\jpsi\rho^0$
yield decreased by 23\% and that for $\jpsi\pipi$ (non-resonant) increased by
27\%, the total $\jpsi\pipi$ yield increased by less than 1\%.  This check is
also equivalent to varying the parametrization for $\jpsi\pipi$ (non-resonant)
and is strictly a test of the stability of the total $\jpsi\pipi$ yield.  It
should be noted that the simple Breit-Wigner is technically the wrong shape
to use for the $\rho^0$ parametrization.

Another way to model the backgrounds is to use a smoothing algorithm on the
data rather than impose definite PDF shapes.  The chosen method creates a 
Gaussian kernel for each event in order to build a shape that describes the 
input data~\cite{Keys}.  The resulting PDF follows 
fluctuations in the input data as accurately as possible and checks how
strongly the fitted signal yield depends on the chosen method of background
parameterization.  Changing the background modelling in this way alters the
total fitted yield by less than one event.

The signal yield can also be calculated by counting the number of data
events passing all the cuts and subtracting the estimated numbers of 
background and $\jpsi\KS$ events.  This method gives $91.0 \pm 16.6$ 
$\jpsi\pi^+\pi^-$ events, to be compared to the $89.9 \pm 13.3$ given by the
fit to the $M(\pipi)$ distribution.

There are $43.3 \pm 12.7$ events in the $\rho^0$ resonance component.  On the 
assumption that the non-$\rho^0$ signal component is well modelled by 
$\jpsi\pipi$ phase space, we measure 
$\mathcal{B}(\Bz\to\jpsi\rho^0) = (2.4 \pm 0.7)\times 10^{-5}$, where the 
error is statistical only.

\section{Conclusion}
In summary, we have measured the branching fraction for \Bz\ meson decay to the
final state $\jpsi\pipi$ to be $(5.0 \pm 0.7 \,({\rm stat}) \pm 0.6 
\,({\rm syst}))\times 10^{-5}$.  The technique of fitting the mass spectrum of 
the $\pi\pi$ system will become important for future measurements 
in this channel, particularly as it will provide separation between the 
resonant and non-resonant components given a larger data set.  It is 
important to reliably isolate the $\jpsi\rho^0$ component in order to
contribute to a measurement of the Unitarity Triangle angle $\beta$.

\section{Acknowledgments}
\label{sec:Acknowledgments}
\input acknowledgements

\end{document}

%% file: authors_win2002.tex
\begin{center}
\small

The \babar\ Collaboration,
\bigskip

B.~Aubert,
D.~Boutigny,
J.-M.~Gaillard,
A.~Hicheur,
Y.~Karyotakis,
J.~P.~Lees,
P.~Robbe,
V.~Tisserand,
A.~Zghiche
\inst{Laboratoire de Physique des Particules, F-74941 Annecy-le-Vieux, France }
A.~Palano,
A.~Pompili
\inst{Universit\`a di Bari, Dipartimento di Fisica and INFN, I-70126 Bari, Italy }
% ...

\end{center}\newpage

%% file: acknowledgements.tex
We are grateful for the 
extraordinary contributions of our \pep2\ colleagues in
achieving the excellent luminosity and machine conditions
that have made this work possible.
The success of this project also relies critically on the 
expertise and dedication of the computing organizations that 
support \babar.
The collaborating institutions wish to thank 
SLAC for its support and the kind hospitality extended to them. 
This work is supported by the
US Department of Energy
and National Science Foundation, the
Natural Sciences and Engineering Research Council (Canada),
Institute of High Energy Physics (China), the
Commissariat \`a l'Energie Atomique and
Institut National de Physique Nucl\'eaire et de Physique des Particules
(France), the
Bundesministerium f\"ur Bildung und Forschung
(Germany), the
Istituto Nazionale di Fisica Nucleare (Italy),
the Research Council of Norway, the
Ministry of Science and Technology of the Russian Federation, and the
Particle Physics and Astronomy Research Council (United Kingdom). 
Individuals have received support from 
the A. P. Sloan Foundation, 
the Research Corporation,
and the Alexander von Humboldt Foundation.

%% file: conf_paper.bbl
\begin{thebibliography}{99}
\bibitem{Bishai:1995yj}
CLEO Collaboration, M.~Bishai {\it et al.},
Phys.\ Lett.\ B {\bf 369}, 186 (1996).\\
Eq.~1 is used to estimate both the Standard Model prediction for 
$\mathcal{B}(\Bz\to\jpsi\rho^0)$ and that for $\mathcal{B}(\Bz\to\jpsi\pipi)$ 
(non-resonant), where the factor of 0.5 is excluded from the latter 
calculation.  The sum of the two results gives the prediction for
$\mathcal{B}(\Bz\to\jpsi\pipi)$.
\end{thebibliography}
